\documentclass[final,5p,times]{elsarticle}
\usepackage{graphicx}
\usepackage{amssymb}
\usepackage{mathtools, cuted}
\usepackage{physics}
\usepackage{siunitx}
\usepackage{hyperref}
\usepackage{cleveref}
\usepackage[utf8]{inputenc}
\usepackage{bbm}
\usepackage{amsmath}
\usepackage{url}
\usepackage{bbm}
\usepackage{xcolor}
\usepackage{subfig}
\begin{document}

\begin{frontmatter}


\title{Light emission properties in a double quantum dot molecule immersed in a cavity: phonon-assisted tunneling}



\author[un]{Vladimir Vargas-Calderón}
\ead{vvargasc@unal.edu.co}
\author[un]{Herbert Vinck-Posada}
\ead{hvinckp@unal.edu.co}
\address[un]{Grupo de Superconductividad y Nanotecnología, Departamento de Física, Universidad Nacional de Colombia, AA 055051, Bogotá, Colombia}

\begin{abstract}
Two main mechanisms dictate the tunneling process in a double quantum dot: overlap of excited wave functions, effectively described as a tunneling rate, and phonon-assisted tunneling. In this paper, we study different regimes of tunneling that arise from the competition between these two mechanisms in a double quantum dot molecule immersed in a unimodal optical cavity. We show how such regimes affect the mean number of excitations in each quantum dot and in the cavity, the spectroscopic resolution and emission peaks of the photoluminescence spectrum, and the second-order coherence function which is an indicator of the quantumness of emitted light from the cavity.
\end{abstract}


\end{frontmatter}

%\linenumbers

\section{Introduction}
\label{S:1}

One of the most promising physical systems for the implementation of quantum technologies, particularly quantum computing, are solid state quantum dots (QDs). These versatile devices have been placed in the core of physical research in past years because they are a playground to study fundamental quantum physics. They are also useful to propose new technological devices with applications in communications, life sciences, metrology, among others. One of the main interests of physicists and engineers on QDs arises from the idea that current silicon-based lithographic techniques from the chip industry would allow the exploitation of QDs scalability. Therefore, the study of the building blocks of many QDs arrays is necessary. The simplest, but still not fully understood of these blocks, is the double QD molecule, which is a physical system where two QD-based qubits interact with each other. Coherent control of tunneling in a double QD is essential for the operation of quantum information technologies based on QD qubits \cite{divincenzo1998,divincenso1999}.

However, it is known that inter-dot tunneling might be incoherently assisted by phonons of the QDs' surrounding lattice, causing undesired tunneling \cite{brandes1999,Fujisawa932}. The physical mechanism of this assistance, called the phonon-assisted tunneling (PhAT), is the Coulomb interaction between an excitation that is transferred from one QD to the other one, and the ion lattice. Mainly, when there is an energy mismatch between the QDs, the excitation tunneling implies the emission or absorption of a phonon that compensates for this energy \cite{VARGASCALDERON2019}. Therefore, bath engineering plays a central role in controlling the dynamics of double QDs, which is primarily dominated by two competing mechanisms: coherent tunneling and PhAT, which leads to incoherent tunneling. It is usual to control coherent tunneling using voltage gates to select the characteristic energies of excitations in the QDs \cite{wang2011,Li_2018}. With these gates, it is also possible to use the Coulomb blockade to restrict tunneling from one QD to another one \cite{Najdi_2018}. Moreover, by embedding a double QD in an optical micro-cavity, further control of the state of the system is achievable. It has been known for years that the strong-coupling regime in a QD-cavity system provides the feasibility of quantum information tasks \cite{Hennessy2007,haroche2001}, since the cavity partially protects the localized QD (or double QD) system from decoherence \cite{Mabuchi1372}.

In this paper, we characterize and explore the simultaneous action of both coherent and incoherent inter-dot tunneling in some observables of a double QD-cavity system. We pay particular attention to the existence of different inter-dot tunneling regimes, which give rise to likewise different behaviors of the observables. For instance, Ref. \cite{borges2010} defines a weak and a strong tunnelling regime with the boundaries 0.01-0.1 \si{\milli\electronvolt} \cite{tackeuchi2000} and 1-10 \si{\milli\electronvolt} \cite{ortner2005,emary2007}, respectively. As it is expected, these regimes depend upon the other Hamiltonian and dissipative parameters that describe the dynamics of the system.

This paper is organized as follows. In \cref{sec:theory}, we lay out the theoretical tools to study the competition between coherent and incoherent tunneling in a double QD-cavity system. \Cref{sec:results} shows the main results based on our numerical calculations. Finally, we present conclusions in \cref{sec:conclusions}.

\section{Theoretical Framework \label{sec:theory}} 

We study a double QD coupled through inter-dot tunnelling~\cite{jefferson1996} interacting with a confined electromagnetic single-mode of a microcavity. Each QD is considered as a two-level system, so that a QD may be in the ground state or an excited state. In general, this excited state can be defined by the trapping of an electric charge in an electric gate defined QD, or an exciton state of a semiconductor QD, or any of the multiple variants that are experimentally available to build a qubit out of a QD. The interaction between each QD and the cavity mode is taken to be described by the dipole approximation in the rotating wave approximation, which is outlined by the Jaynes-Cummings interaction~\cite{jaynes-cummings1963}. Hereafter, we follow the construction of our previous work \cite{VARGASCALDERON2019}. Thus, the Hamiltonian of the double QD-cavity system is
\begin{align}
\begin{aligned}
H =& \sum_{j=1}^2\hbar\omega_j\sigma_j^\dagger\sigma_j + \hbar\omega_0a^\dagger a + \hbar T (\sigma_1^\dagger\sigma_2 + \sigma_1\sigma_2^\dagger)\\
&+ \sum_{j=1}^2 \hbar g_j(\sigma_j^\dagger a + \sigma_j a^\dagger)
\end{aligned}
\end{align}
where  $\sigma_j^\dagger (\sigma_j)$ is the creation (annihilation) operator for the $j$-th two-level QD with frequency $\omega_j$, $a^\dagger (a)$ is the creation (annihilation) operator for the cavity mode with frequency $\omega_0$, $T$ is the tunnelling rate and $g_j$ is the QD-cavity interaction constant between the mode and the $j$-th QD. However, the physical system entangles with the surrounding environment, leading to decoherence processes that are well-described in the Born-Markov approximation by the quantum master equation~\cite{breuer2007,VARGASCALDERON2019}:
\begin{align}
\begin{aligned}
\dot{\rho}(t) =& -\frac{i}{\hbar}[H, \rho(t)] + \sum_{j=1}^2(\gamma_j \mathcal{D}_{\sigma_j} + P_j\mathcal{D}_{\sigma_j^\dagger})   +  P\mathcal{D}_{a^\dagger} \\
&+ \kappa \mathcal{D}_{a} + \gamma_T \mathcal{D}_{\sigma_1^\dagger\sigma_2 } + P_T \mathcal{D}_{\sigma_1\sigma_2^\dagger},
\end{aligned}\label{eq:qme}
\end{align}
where $\mathcal{D}_O =  O\rho(t)O^\dagger - \frac{1}{2}\{O^\dagger O, \rho(t)\}$ are the dissipative superoperators in Lindblad form corresponding to each decay channel. In particular, the $j$-th QD couples to 
external excitation reservoirs which incoherently pump the QD at a rate $P_j$, and couples to leaky photonic modes that deplete the QD from excitations via spontaneous emission at a rate $\gamma_j$~\cite{tejedor2004}. Additionally, due to imperfections in the cavity mirrors, photons of the cavity escape at a rate $\kappa$. There has also been observed incoherent cavity pumping (whose rate is denoted as $P$), which is attributed to non-resonant QDs inside semiconductor cavities \cite{winger2009,laucht2010}. The last two terms of \cref{eq:qme} correspond to PhAT \cite{bagheri2012,rozbicki2008,karwat2011,Karwat2014,VARGASCALDERON2019}, which are similar in form to cavity phonon-feeding~\cite{majumdar2011}. Here, $\gamma_T$ is the rate associated to incoherent tunnelling from QD2 to QD1, and the converse for $P_T$. In this description of the system, we neglect pure dephasing (because it is negligible at low temperatures), as well as the cavity phonon-feeding, in order to isolate the tunnelling mechanisms in the description of observables such as the expected number of excitations in the electromagnetic mode of the cavity, in the QDs, the photoluminiscence (PL) spectrum and the $g^{(2)}$ second-order coherent function.

The expected number of excitations are easily obtained by solving the kernel of the Liouvillian, i.e. $\dot{\rho} = \mathcal{L}\rho(t) = 0$, which has a steady state solution $\rho_{\text{ss}}$. Thus, the expected number of excitations are $\expval{\sigma_1^\dagger\sigma_1},\expval{\sigma_2^\dagger\sigma_2}$ and $\expval{a^\dagger a}$ for the QD1, QD2 and cavity, respectively, where the expected values are taken with respect to $\rho_\text{ss}$.

The PL spectrum can be obtained \cite{scully_zubairy_1997,tejedor2004,VARGASCALDERON2019} through the Wiener-Khintchine theorem, which relates the power intensity $I(\omega)$ to the two-point correlation function of the mode operators as
\begin{align}
I(\omega) \propto \frac{\kappa}{\pi}\lim_{t\to\infty}\int_{0}^\infty \expval{a^\dagger(t)a(t+\tau)}e^{i\omega\tau} d\tau.\label{eq:spectrum}
\end{align}
The evaluation of this correlation function can be done using the quantum regression theorem~\cite{breuer2007}. The PL spectrum gives information of the optical transitions that occur in the system. Particularly, only decay transitions between consecutive excitation manifolds will contribute to the PL spectrum. In the regime of low incoherent pumpings, only the lower excitation manifolds are occupied, so that it is expected that the only optical transitions occur between the 0th and 1st excitation manifolds. To study such transitions, we adopt Refs. \cite{torres2014,santiEcheverri2018}, where a thorough examination of the transitions of an open quantum system is given. By neglecting gain processes, the Liouvillian decouples into independent blocks, whose complex eigenvalues contain the information of the transition energies, as well as their corresponding linewidths. As discussed before, we are only interested in transitions between the 0th and 1st excitation manifolds, spanned by the bare states $\ket{0 G G}$ and $\{\ket{1GG}, \ket{0XG}, \ket{0GX}\}$, respectively, where $\ket{n \substack{G\\X} \substack{G\\X}} := \ket{n}\otimes \ket{\substack{G\\X}} \otimes \ket{\substack{G\\X}}$ is the state of the QD-cavity system. Here, $\ket{n}$ is the Fock state of the electromagnetic mode in the cavity and $\ket{\substack{G\\X}}$ denotes the $\substack{\text{ground}\\ \text{excited}}$ state of either the first or the second QD. As shown in \cite{torres2014}, the spectrum of the Liouvillian block that corresponds to transitions between the 0th and 1st excitation manifolds is given by the eigenvalues of the matrix
\begin{strip}
\begin{equation}
    \mathcal{M}^{(0,1)} = \frac{1}{i}\left(K^{(1)}\otimes \mathbbm{1}_1 - \mathbbm{1}_3\otimes K^{*(0)}\right) +\sum_{s=1}^2\frac{\xi_s}{2}\left(2C_s^{(1)}\otimes C_s^{*(0)} - [C_s^\dagger C_s]^{(1)}\otimes \mathbbm{1}_1-\mathbbm{1}_3\otimes [C_s^\dagger C_s]^{\intercal(0)}\right),
\end{equation}
\end{strip}
where $\xi_1 = \gamma_T, \xi_2=P_T, C_1=\sigma_1^\dagger\sigma_2$, and $C_2=\sigma_2^\dagger\sigma_1$. Also, $K = H - i\hbar (\frac{\gamma_1}{2}\sigma_1^\dagger \sigma_1 + \frac{\gamma_2}{2}\sigma_2^\dagger\sigma_2 + \frac{\kappa}{2}a^\dagger a)$. Moreover, the projections of any operator $O$ onto the 0th and 1st excitation manifolds are
\begin{strip}
\begin{equation}
    O^{(0)} = \bra{0GG}O\ket{0GG}, \quad
    O^{(1)} = \begin{pmatrix}
    \bra{1GG} O \ket{1GG} & \bra{1GG} O \ket{0XG} & \bra{1GG} O \ket{0GX}\\
    \bra{0XG} O \ket{1GG} & \bra{0XG} O \ket{0XG} & \bra{0XG} O \ket{0GX}\\
    \bra{0GX} O \ket{1GG} & \bra{0GX} O \ket{0XG} & \bra{0GX} O \ket{0GX}
    \end{pmatrix}.
\end{equation}
\end{strip}
Thus, the explicit matrix $\mathcal{M}^{(0,1)}$ is given by
\begin{strip}
\begin{equation}
    \mathcal{M}^{(0,1)} = \frac{1}{i}\begin{pmatrix}
    \omega_0 - i \kappa/2   &   g_1 &   g_2\\
    g_1 & \omega_1  - i\gamma_1/2 - iP_T/2    &   T\\
    g_2 &   T   &   \omega_2 - i\gamma_2/2 - i\gamma_T/2
    \end{pmatrix},\label{eq:matrix}
\end{equation}
\end{strip}
and its characteristic polynomial gives analytic expressions for the eigenvalues. Their imaginary part are the transition frequencies between the 0th and 1st excitation manifolds, whereas their real part are the corresponding half widths at half maximum \cite{VARGASCALDERON2019}.

Finally, the $g^{(2)}$ second-order coherence function is defined as
\begin{align}
    g^{(2)}(\tau) = \frac{\expval{a^\dagger(0) a^\dagger(\tau) a (\tau) a(0)}}{\expval{a^\dagger(0)a(0)}}.
\end{align}

\section{Results and Discussion\label{sec:results}}

\begin{figure*}[!t]
    \centering
    \includegraphics[width=\textwidth]{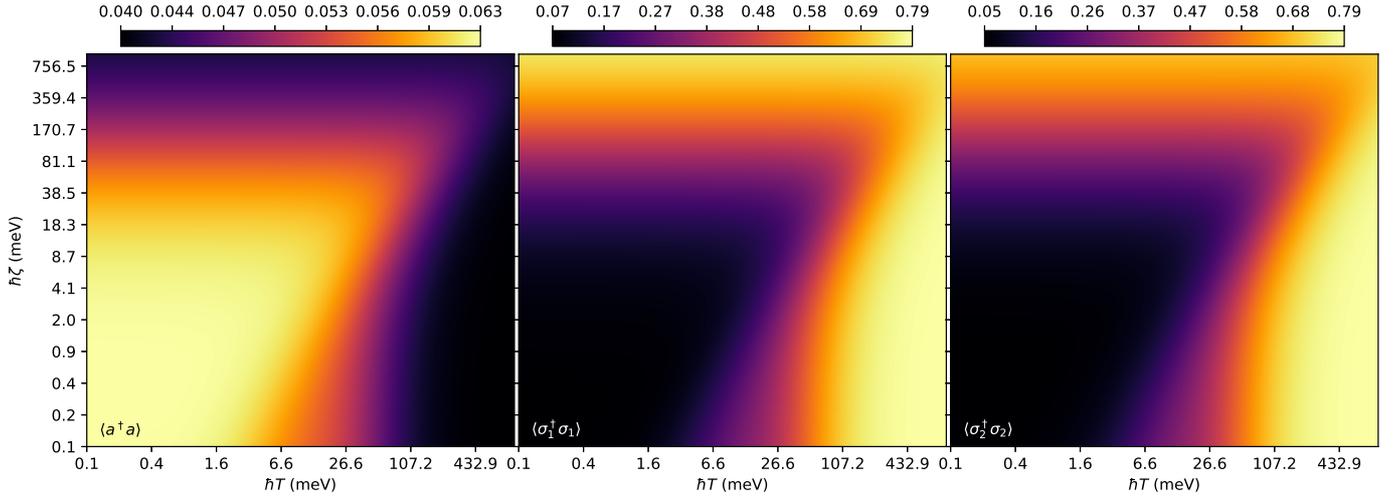}
    \caption{Log-log plots of the expected excitations of the cavity (left panel), QD1 (centre panel) and QD2 (right panel) in the steady state $\rho_\text{ss}$ as a function of $T$ and $\zeta$. \label{fig:expected}}
\end{figure*}

For our numerical calculations, we consider the estimated Hamiltonian and dissipative parameters found in Ref.~\cite{laucht2010_2}, where two self-assembled InGaAs QDs are strongly coupled to a single electromagnetic mode confined in a cavity. Here, the cavity mode energy is approximately $\hbar\omega_0 = 1.218~\si{\electronvolt}$. The exciton energies $\hbar\omega_1$ and $\hbar\omega_2$ for the 1st and 2nd QDs, respectively, can be tuned with an applied bias voltage. In our study, we consider one QD to be resonant with the cavity ($\hbar\omega_1=\hbar\omega_0$), and the other one to be slightly off-resonant ($\hbar\omega_2 = \hbar\omega_1 + 0.1~\si{\milli\electronvolt}$), because we want to study how this energy mismatch gives room to acoustic phonons to assist the inter-dot tunneling. The near-resonant condition exhibits the relevant physical behavior of the system's optical properties, such as the PL emission peaks coalescence, as will be seen later. Additionally, the QD-cavity coupling constants are $\hbar g_1=44~\si{\micro\electronvolt},\hbar g_2 = 51~\si{\micro\electronvolt}$. Regarding the dissipative parameters, the excitonic pumping to the QDs are $\hbar P_1=1.5~\si{\micro\electronvolt}, \hbar P_2=1.9~\si{\micro\electronvolt}$, the spontaneous emission of the QDs are $\hbar\gamma_1=0.1~\si{\micro\electronvolt}, \hbar\gamma_2 = 0.8~\si{\micro\electronvolt}$, the escape of photons from the cavity is $\hbar\kappa=147~\si{\micro\electronvolt}$, and the pumping due to non-resonant QDs in the cavity to the cavity is $\hbar P = 5.7~\si{\micro\electronvolt}$. We remark that the rate $\kappa$ is larger than the QD-cavity coupling constants. As a theoretical assumption, from now on we take $\hbar g_1=440~\si{\micro\electronvolt},\hbar g_2 = 510~\si{\micro\electronvolt}$ so that the Jaynes-Cummings interaction is clearly dominant over the escape of photons, guaranteeing strong coupling in our calculations. 

Although the authors of Ref. \cite{laucht2010_2} do not report the temperature, a common value in these sorts of experiments is around 4 \si{\kelvin}~\cite{winger2008hennessy}. Similarly as in the case of cavity phonon-feeding \cite{majumdar2011}, and also as in the case of thermal baths~\cite{breuer2007}, the PhAT rates are related through $\gamma_T = (n_\text{th}+1)\zeta$ and $P_T = n_\text{th}\zeta$, when QD1 is red detuned with respect to QD2, where $n_\text{th}(\Delta, T) = 1/(\exp[\hbar\Delta/k_B T]-1)$, being $\Delta = \omega_2-\omega_1$, and $T$ the temperature, not the tunnelling rate. In our case, $n_\text{th}\approx 3$. \Cref{fig:expected} shows the expected excitations for the cavity, QD1 and QD2 in the steady state of the system as a function of $T$ and $\zeta$. The expected number of photons in the cavity is very low at every value of $T$ and $\zeta$ due to the magnitude of the photon escape rate from the cavity. At low (high) values of the tunneling and PhAT rates, the number of photons is larger (smaller), which is the opposite behavior of what occurs in the QDs expected number of excitations. It is common to all three subsystems that there are two very different regimes. High tunneling rates delocalize the excitations in the QDs, diminishing Rabi oscillations with the electric field mode. Since QD incoherent pumping is larger than spontaneous emission, in the steady state, QDs will have a greater number of excitations. Note that this effect occurs at PhAT rates 1--2 orders of magnitude larger than the tunneling rates needed for this delocalization, which is characteristic of a molecular state. It is worth noting that higher temperature leads to larger $n_\text{th}$, meaning that lower PhAT rates are needed for this delocalization. This can be seen from the fact that both $\gamma_T$ and $P_T$ are directly proportional to $n_\text{th}$, and increasing $\zeta$ is similar--but not equal to--increasing the temperature, since high temperatures ensure $\gamma_T\approx P_T$. The opposite occurs with low temperatures, where assisted tunneling by phonons is enhanced in the direction from the blue-detuned QD to the red-detuned QD.

\begin{figure*}[!ht]
    \centering
    \includegraphics[width=\textwidth]{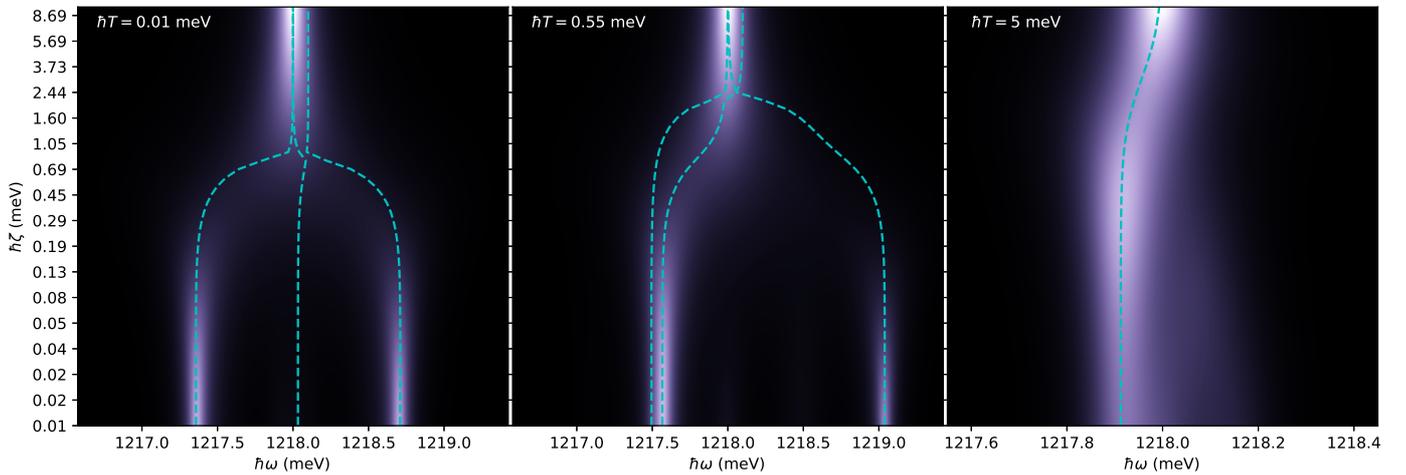}
    \caption{PL spectra for several values of PhAT $\zeta$ rate for three different values of the tunnelling rate $T$: 0.01 \si{\milli\electronvolt} (left panel), 0.55 \si{\milli\electronvolt} (centre panel) and 5 \si{\milli\electronvolt} (right panel). Cyan dashed lines are the allowed optical transitions between the 0th and 1st excitation manifolds, which are the imaginary part of the eigenvalues of the matrix in \cref{eq:matrix}. Peaks of the PL spectra are shown in light colour.
    \label{fig:spectra}}
\end{figure*}

\begin{figure*}[!ht]%
    \centering
    \subfloat[]{{\includegraphics[width=0.49\textwidth]{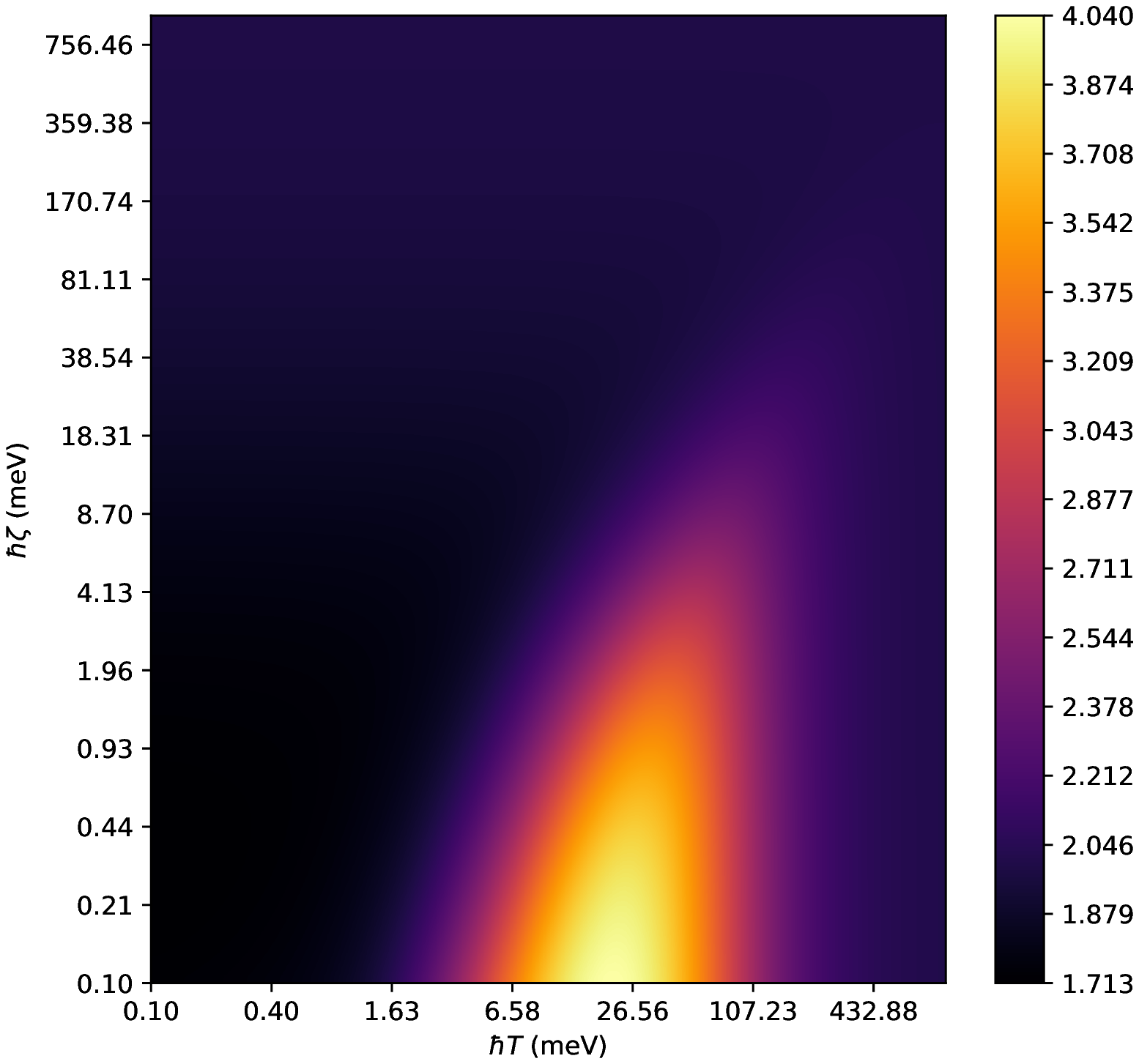} }}%
    \subfloat[]{{\includegraphics[width=0.49\textwidth]{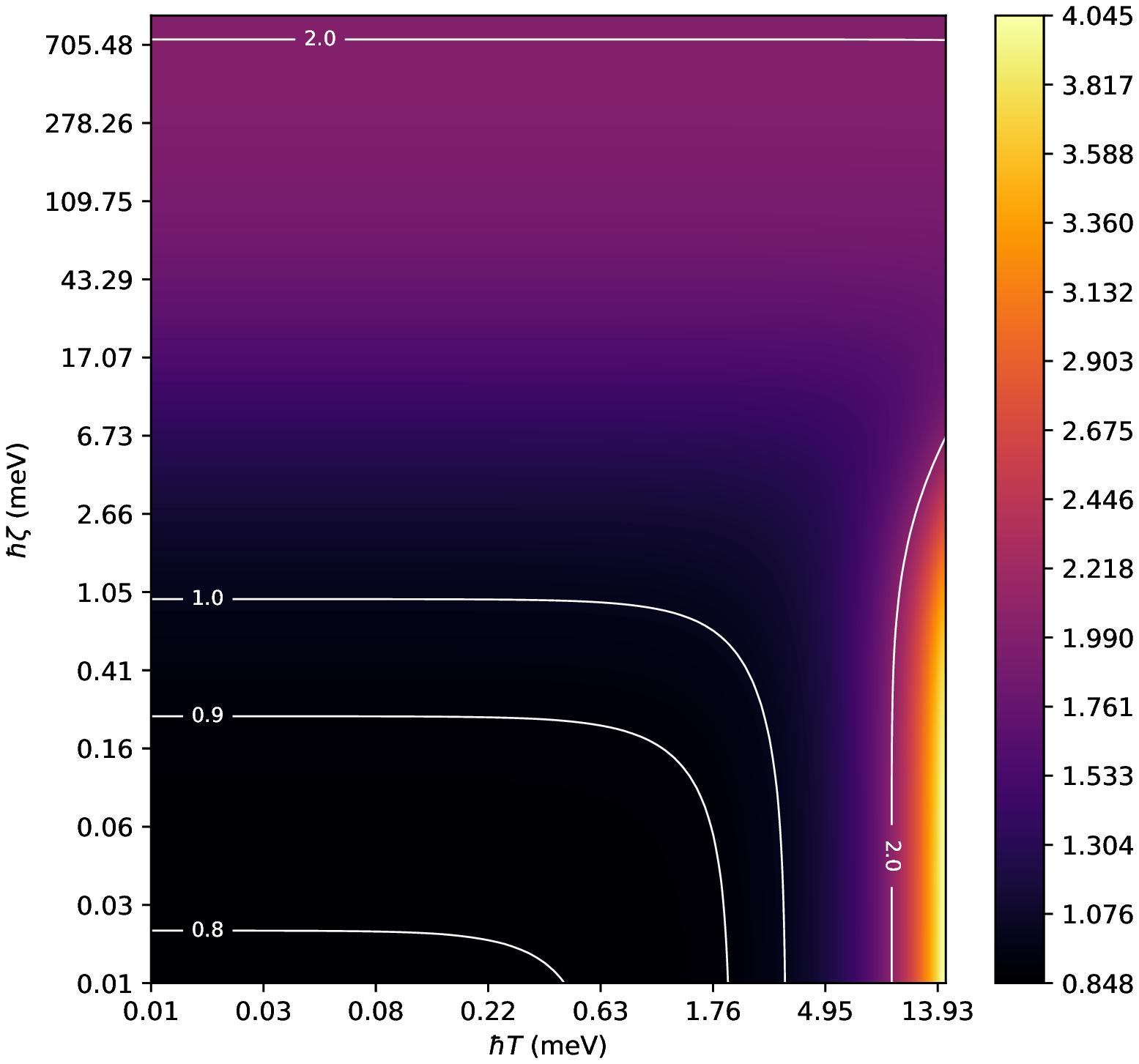} }}%
    \caption{Log-log plot of $g^{(2)}(\tau=0)$ as a function of $T$ and $\zeta$ using the parameters $\hbar P_1=1.5~\si{\micro\electronvolt}, \hbar P_2=1.9~\si{\micro\electronvolt}, \hbar\gamma_1=0.1~\si{\micro\electronvolt}, \hbar\gamma_2 = 0.8~\si{\micro\electronvolt}, \hbar\kappa=147~\si{\micro\electronvolt}, \hbar P = 5.7~\si{\micro\electronvolt}, \hbar g_1=440~\si{\micro\electronvolt},\hbar g_2 = 510~\si{\micro\electronvolt}$ in the left panel and $\hbar P_1=\hbar P_2 = 1.752~\si{\milli\electronvolt}, \hbar\gamma_1=\hbar \gamma_2 = 17.52~\si{\micro\electronvolt}, \hbar\kappa=12.18~\si{\milli\electronvolt}, \hbar P = 0\si{\micro\electronvolt}, \hbar g_1= \hbar g_2 = 1.218~\si{\milli \electronvolt}$ in the right panel.}%
    \label{fig:g2}%
\end{figure*}

A fairly common measurement in these experiments, which shows signatures of strong coupling, is the PL spectrum. In \cref{fig:spectra} PL spectra are shown for three different values of the tunneling rate $T$, as well as for a continuous sweep of the PhAT $\zeta$ rate in each of the values of $T$. The three optical transitions from the 1st to 0th excitation manifolds are shown in each panel in dashed lines. The left and central panels show a coalescence behavior, caused by a dynamical phase transition where spectroscopic resolution is lost due to the action of PhAT~\cite{VARGASCALDERON2019}. For low tunneling rates, the lateral transitions show luminescence, while the central transition remains inactive, as shown in the left panel. As the tunneling rate increases, the central transition redshifts, and the lateral transitions blue shift, so that the left and central transitions come very close to each other, as shown in the central panel. As $T$ continues to increase, the left and central transitions eventually change places. The right transition fades out, leaving just one peak, corresponding to the initially left transition, shown in the right panel. One could think that coherent tunneling leads to a loss in spectroscopic resolution, just as PhAT does. However, this is false, since an increase in $T$ does not provoke coalescence, where transition peaks merge in the so-called exceptional points of the dynamical phase transitions. Furthermore, in the case of $\hbar T = 5~\si{\milli\electronvolt}$ there is also coalescence at larger values of $\zeta$ (not shown in the plot). Finally, it is important to highlight that we have swept a large energy range of electron-phonon coupling $\hbar\zeta$ in the analysis of mean number of excitations and $g^{(2)}$. It must be noted that at high values of $\zeta$, the assumption of a single excitation level in each QD becomes weak, and the behavior described in \cref{fig:expected,fig:g2} can only be confirmed through a more exact treatment such as a multiexcitonic quantum dot \citep{VINCK200699}, which falls beyond the scope of this work. However, it has been recently experimentally observed that the consideration of a single exciton level in a DQD-cavity system can explain its dynamics at relatively high electron-phonon couplings with respect to the cavity energy (around 38\%, which is much higher than the other interaction parameters \cite{petta2018}).

\section{Conclusions\label{sec:conclusions}}

We studied the combined effect of coherent and dissipative or incoherent tunneling between two coupled QDs immersed in a unimodal cavity. This study was directed towards the analysis of different tunneling regimes that are physically observable in the mean number of excitations in each part of the double QD-cavity system, in the PL spectrum and the second-order coherence function $g^{(2)}$.

In the mean number of excitations, coherent tunneling and PhAT showed a similar effect, where a delocalization of the excitations stored in the QDs occurs. However, the analysis of PL spectra showed a clear indicator that differentiates PhAT from coherent tunneling, where the former leads to emission peaks coalescence (characteristic of a dynamical phase transition) and the latter to a shift of the peaks at fixed PhAT rate.
Interestingly, for the set of parameters considered in the analysis that allowed us to draw the conclusions mentioned above, the second-order coherence function shows a non-monotonic dependence on $T$ and highlights a bounded region of very classical light. Why this happens remains an open question. However, a search in the parameter space revealed many regions where anti-bunched light can be produced. In general low values of coherent tunnelling rates and PhAT rates allow $g^{(2)}(\tau=0) < 1$.

\section*{Acknowledgements}
The authors acknowledge partial financial support from COLCIENCIAS under the project ``Emisión en sistemas de Qubits Superconductores acoplados a la radiación. Código 110171249692, CT 293-2016, HERMES 31361'', and project ``Control dinámico de la emisión en sistemas de qubits acoplados con cavidades no-estacionarias, HERMES 41611''.

\section*{Data availability}
The raw data required to reproduce these findings are available to download from \cite{dataset}. The processed data required to reproduce these findings are available to download from \cite{dataset}.

\bibliographystyle{model1-num-names}
\bibliography{sample.bib}






% \end{thebibliography}

\end{document}